\documentclass[twocolumn,longbibliography]{revtex4-1}
\usepackage{amsmath}
\usepackage{amssymb}
\usepackage{bm}
\usepackage{epsfig}
\usepackage{graphicx}
\usepackage{color}
\usepackage[colorlinks]{hyperref}
\usepackage[english]{babel}

\usepackage[bbgreekl]{mathbbol}

\begin{document}
\newcount\timehh  \newcount\timemm
\timehh=\time \divide\timehh by 60
\timemm=\time
\count255=\timehh\multiply\count255 by -60 \advance\timemm by \count255

\title{Quantum interference effect on exciton transport in monolayer semiconductors}

\author{M.M. Glazov}

\address{Ioffe Institute, 194021 St.-Petersburg, Russia}

\begin{abstract}
We study theoretically weak localization of excitons in atomically-thin transition metal dichalcogenides. The constructive interference of excitonic de Broglie waves  on the trajectories forming closed loops results in a decrease of the exciton diffusion coefficient. We calculate the interference contribution to the diffusion coefficient for the experimentally relevant situation of exciton scattering by acoustic phonons and static disorder. For the acoustic phonon scattering, the quantum interference becomes more and more important with increasing the temperature. Our estimates show that the quantum contribution to the diffusion coefficient is considerable for the state-of-the-art monolayer and bilayer transition metal dichalcogenides. 
\end{abstract}

\maketitle

\emph{Introduction}.
Two-dimensional (2D) semiconductors based on transition metal dichalcogenides (TMDC) form a versatile platform for excitonic physics~\cite{RevModPhys.90.021001}. Direct band gap~\cite{Splendiani:2010a,Mak:2010bh} and chiral selection rules~\cite{Xiao:2012cr} in these atomically thin semiconductors provide direct access to the temporal and spatial dynamics of Coulomb interaction bound electron-hole pairs. Optical experiments can visualize the exciton transport and 
demonstrate their propagation in monolayer (ML)~\cite{Mouri:2014a,Kato2016,doi:10.1021/acs.jpclett.7b00885,Cadiz:2018aa,PhysRevLett.120.207401,Wang:2019aa,leon2019hot,zipfel2019exciton} and heterobilayer transition metal dichalogenides~\cite{Rivera688,Unuchek:2018aa}.

The key quantity in transport phenomena is the diffusion coefficient $D$ which describes the propagation of the particles on the macroscopic, i.e., larger than the mean free path, distances. The diffusion coefficient is determined by the velocity autocorrelation functions and, according to the fluctuation-dissipation theorem, can be related to the conductivity or mobility of the particles. In the quasi-classical approach the particles can be represented as wavepackets which move freely and experience relatively rare scattering events. Thus, $D$ in a 2D system can be written as~\cite{afonin:1985a,arseev98,durnev:wl} 
\begin{equation}
\label{D}
D= \left\langle \frac{v^2 \tau}{2} \right\rangle.
\end{equation}
Here the angular brackets denote the averaging over the thermal distribution, $v$ is the particle velocity and $\tau$ is the momentum scattering time. The diffusion coefficient is also of prime importance for describing the nonlinear propagation of excitons and the transport driven by the thermal gradients and phonon fluxes both in TMDC MLs~\cite{PhysRevLett.120.207401,PhysRevB.100.045426,Perea-Causin:2019aa} and conventional semiconductors~\cite{BENSON1997313,PSSB:PSSB45,Ivanov_2002,PhysRevLett.92.117404}.

In TMDC MLs the exciton thermalization is very efficient and in few picoseconds after photoexcitation they form a non-degenerate Boltzmann gas~\cite{Selig_2018}. The quasi-classical description of transport effects holds at~\cite{ll10_eng,afonin:1985}
\begin{equation}
\label{IR}
\frac{k_B T \tau}{\hbar} \gg 1,
\end{equation}
where $k_B$ is the Boltzmann constant, $T$ is the temperature. This condition means that the collisional broadening of the particle energy levels, $\sim \hbar/\tau$, is by far inferior than the characteristic energy of the particle, $\sim k_B T$. The condition~\eqref{IR} can be also recast in the form of Ioffe-Regel criterion $l/\lambda \gg 1$, where $\lambda$ is the exciton de Broglie wavelength and $l$ is its mean free path. 

Recent experiments and theoretical analysis demonstrate that the criterion~\eqref{IR} can be easily violated for excitons in TMDC MLs~\cite{Selig:2016aa,PhysRevLett.119.187402,shree2018exciton}. In these materials the interaction of excitons with phonons in enhanced as compared to conventional quasi-2D systems such as semiconductor quantum wells due to higher density of states of 2D phonons~\cite{shree2018exciton}. The exciton momentum relaxation time by acoustic phonons with linear dispersion at not too low temperatures ($k_B T \gg Ms^2$) can be presented as~\cite{PhysRevB.85.115317,shree2018exciton,supp}
\begin{equation}
\label{tau:ac}
\tau = \frac{Ms^2}{k_B T} \tau_0, 
\end{equation}
where $M$ is the exciton translational mass, $s$ is the speed of sound, and $\tau_0$ is a constant related to the strength of the exciton-phonon interaction. The exciton scattering time by acoustic phonons $\propto T^{-1}$ because in the quasi-elastic regime the number of available phonons scales linearly with the temperature. The diffusion coefficient~\eqref{D} in a quasi-classical limit acquires the form
\begin{equation}
\label{D:ac}
D= \frac{k_B T \tau}{M} = s^2 \tau_0,
\end{equation}
and is temperature independent. The product $k_B T \tau/\hbar = Ms^2 \tau_0/\hbar$ is also temperature independent and on the order of unity in TMDC MLs meaning that the criterion~\eqref{IR} does not, strictly speaking, hold. It is also experimentally confirmed by the thermal broadening of exciton resonance~\cite{Selig:2016aa,shree2018exciton,C9NR04211C}. Furthermore, at the room temperature the exciton interaction with the optical and zone-edge phonons~\cite{Song:2013uq,PhysRevLett.115.115502,PhysRevB.100.041301,He:2020aa} becomes significant which reduces $\tau$ and violates the condition~\eqref{IR} even stronger. All this questions the validity of the quasiclassical description of the exciton transport in 2D semiconductors. 

The analysis of the general case where the product $k_B T \tau/\hbar$ can be on the order of unity is extremely involved~\cite{aa,RevModPhys.80.1355}. Thus, it is important to study the role of the main quantum effects on the exciton transport. Here we calculate the leading order correction to the quasiclassical value of the diffusion coefficient~\eqref{D:ac}. It results from the interference of exciton de Broglie waves on the closed classical trajectories where the phase difference for the clock- and counterclock-wise propagation are almost the same. This effect gives rise to the coherent backscattering of excitons~\cite{1977ZhETF..72.2230I,PhysRevB.77.165341}, and provides the key contribution to the quantum correction to the diffusion coefficient~\cite{Gorkov:WL,aa,bergmann84,arseev98,afonin:1985,afonin:1985a}. The weak localization is usually studied in electronic systems at reduced temperatures by the conductivity measurements~\cite{bergmann84,aa,PhysRevB.99.115414} with very few exceptions including non-degenerate electron gases~\cite{PhysRevLett.58.2106,1988JETPL..47..259D} and quite recently optically~\cite{PhysRevX.8.031021}. 

Here we show that the weak localization can be observed for non-degenerate (Boltzmann) excitons in TMDC mono- and bilayers in diffusion experiments similar to~\cite{Cadiz:2018aa,PhysRevLett.120.207401,Wang:2019aa,zipfel2019exciton}. The calculations are presented for the case of excitons interacting with acoustic phonons. We show that the quantum effects are more pronounced with increasing the temperature due to an interplay of the phonon-induced momentum relaxation and dephasing. We also discuss the effects of other scattering processes in the system. We show that in the most of to date experiments on exciton propagation in TMDC-based systems the  quantum effects are of high importance.

\emph{Qualitative analysis}. 
For the exciton-acoustic phonon interaction at 
\begin{equation}
\label{cond}
 T \gg \frac{Ms^2}{k_B} \sim 1~\mbox{K}
\end{equation}
the transferred energy at a single scattering event $\Delta \varepsilon \sim \sqrt{k_B T Ms^2} \ll k_B T$ and the exciton-phonon scattering is quasi-elastic~\cite{gantmakher87}. For the same reason, the effective potential field created by the phonons and experienced by the excitons is almost static. Thus the excitons can be considered as good quasi-particles and one can use a perturbation theory in $\hbar/(k_B T \tau)$ to calculate the corrections to the classical value~\eqref{D:ac} of the diffusion coefficient.

\begin{figure}[t]
\includegraphics[width=\linewidth]{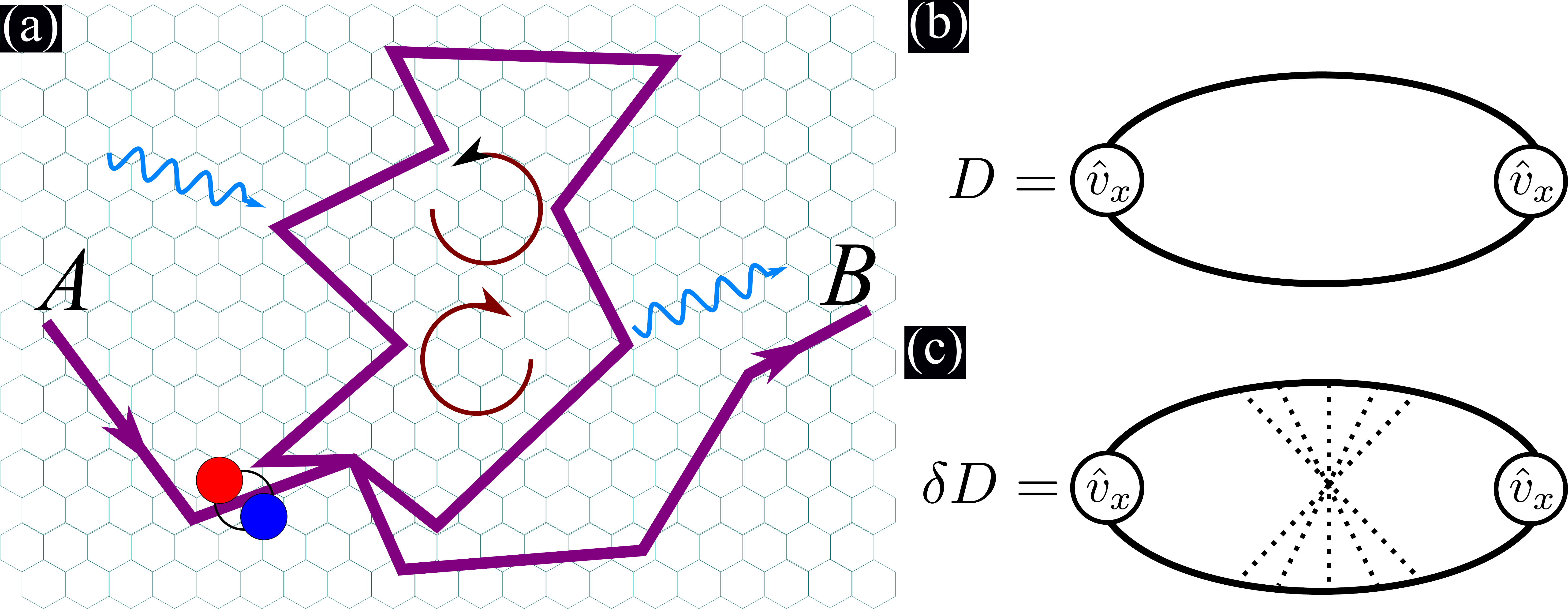}
\caption{\textbf{Illustration of the exciton transport.} (a) Diffusive trajectory of the exciton (shown by blue and red circles) from the point $A$ to the point $B$ with a loop where the exciton can propagate clockwise or counter-clockwise. These trajectories constructively interfere.  Wavy arrows indicate phonons. (b) Diagram describing velocity-velocity autocorrelation function [$D$ in Eq.~\eqref{D:ac}] in the classical approximation for the isotropic scattering. (c) Typical maximally crossed diagram describing leading order quantum correction to (b). Dotted lines describe the phonon propagation. The series of such diagrams with arbitrary number of phonon lines should be summed up to {obtain $\delta D$}, Eq.~\eqref{dD:full}.}\label{fig:wl}
\end{figure}

Here we focus on the  leading order in $\hbar/(k_B T \tau)$ quantum correction to the exciton diffusion coefficient which comes from the weak localization effect. 
It is illustrated in Fig.~\ref{fig:wl}(a) where the self-intersecting classical trajectory with a loop is shown. 
The acquired phases at the clock- and counterclock-wise propagation of excitons in the absence of inelastic scattering processes are exactly the same: $\phi_\circlearrowright=\phi_\circlearrowleft$.
   Formally, these two trajectories are related by the time-reversal which ensures the phase conservation. 
   Since the phase difference $\phi_\circlearrowright-\phi_\circlearrowleft=0$, the interference of the forward and backward paths is constructive and the exciton effectively spends more time at the loop. Consequently, it propagates slower. Hence, the diffusion coefficient is reduced as compared to its classical value. This is the weak localization phenonemon being a precursor of the strong (Anderson) localization~\cite{Gorkov:WL,RevModPhys.80.1355}. 

The magnitude of the quantum correction to the diffusion coefficient is related to the probability of an exciton to return to its starting point in the time interval $(t,t+dt)$, which for a diffusive motion can be estimated in 2D as $\lambda v_T dt/(Dt)$, where $v_T=\sqrt{2k_B T/M}$ is the thermal velocity; $Dt$ gives the area covered by exciton during its diffusion and $\lambda v_T dt$ is the area in the vicinity of the initial spot where the interference occurs. Thus, the correction to the diffusion coefficient is negative and reads
\begin{equation}
\label{dD:est}
\frac{\delta D}{D} \sim -\int_\tau^{\tau_\phi} \frac{\lambda v_T  dt}{Dt} \sim -\frac{\hbar}{k_B T\tau} \ln{\left(\frac{\tau_\phi}{\tau} \right)}.
\end{equation}
In estimation~\eqref{dD:est} we have excluded very short time scales $t\lesssim \tau$, where the motion is ballistic, and also very long time scales $t\gtrsim \tau_\phi$ with $\tau_\phi\gg \tau$ being the phase relaxation time related to the inelastic processes and phonon propagation  which make the scattering potential non-static. Eventually, the phase of the exciton wavefunction becomes broken and the interference stops.

We now estimate the phase relaxation time $\tau_\phi$ for the quasi-elastic exciton-phonon scattering. To that end we consider the mean square of the exciton energy variation as a function of time, $\delta \varepsilon^2(t)$. Since the energy is transferred in small portions $\Delta \varepsilon \ll k_B T$ we use the energy diffusion approximation and obtain $\delta \varepsilon^2(t) = (\Delta \varepsilon)^2 t/\tau$. The energy relaxation time $\tau_\varepsilon$ can be estimated from the condition that $\delta \varepsilon^2(\tau_\epsilon) \sim (k_B T)^2$, which yields $\tau_\varepsilon \sim \tau_0\gg \tau$ in Eq.~\eqref{tau:ac}. By contrast, the phase is lost if the energy uncertainty over time $\tau_\phi$ is on the order of the $\hbar/\tau_\phi$ yielding $
\delta \varepsilon^2(\tau_\phi) \sim \left({\hbar}/{\tau_\phi}\right)^2$ or $\tau_\phi \sim \left[{{\hbar^2\tau_0}/{(k_B T)^2}}\right]^{1/3}$~\cite{PhysRevB.36.5663,afonin:1985}. The same estimate for $\tau_\phi$ can be alternatively obtained considering the phase difference $\phi_\circlearrowright-\phi_\circlearrowleft$ due to the propagation of phonons~\cite{golubentsev}. Notably,  $\tau \ll \tau_\phi \ll \tau_\epsilon$.

Equation~\eqref{dD:est} illustrates the intricate effect of the weak localization on the  exciton transport. On one hand, the quantum correction contains a small parameter of the theory, $\hbar/(k_B T \tau)$. On the other hand, the parametrically long phase relaxation provides large temperature dependent logarithm $\ln{(\tau_\phi/\tau)}\gg 1$. Interestingly, with rising the temperature the prefactor in Eq.~\eqref{dD:est} remains the same as $\tau \propto 1/T$, while the logarithm increases due to the decrease of $\tau_\phi$ with increasing the temperature. Thus, unexpectedly, for exciton-acoustic phonon scattering, the quantum interference becomes progressively more important with increasing the temperature. Below we confirm this qualitative analysis by the diagrammatic calculation.

\emph{Diagrammatic calculation}.
The diagrams contributing to the velocity autocorrelation function and, accordingly, to the diffusion coefficient are shown in Fig.~\ref{fig:wl}, panel (b) for the classical contribution and (c) for the quantum correction. Calculating the retarded and advanced Green's functions  of the excitons $G^{R/A}_{\bm p}(\varepsilon)$ we can neglect inelasticity of the exciton-phonon interaction and use the Born approximation with the result
$G^{R/A}_{\bm p}(\varepsilon) =[\varepsilon - {p^2}/{(2M)} \pm \mathrm i {\hbar}/{(2\tilde \tau)}]^{-1}$.
Here $\tilde\tau^{-1} = \tau^{-1} + \tau_r^{-1}$, with $\tau_r$ being the exciton lifetime and  $\tau$ being the momentum relaxation time, Eq.~\eqref{tau:ac}. For TMDC ML
$\tau_0^{-1} = {M^2(\Xi_c - \Xi_v)^2}/{(\rho \hbar^3)}$~\cite{supp,shree2018exciton}, 
where $\Xi_c$ and $\Xi_v$ are the conduction and valence band deformation potentials, and $\rho$ is the mass density.
Here we take into account interaction of the exciton with the longitudinal acoustic mode. For bilayer TMDC the $(\Xi_c - \Xi_v)^2$ is replaced by the appropriate combination of deformation potentials of relevant phonon modes. The evaluation of the loop in Fig.~\ref{fig:wl}(b)  results in Eq.~\eqref{D:ac} for the exciton diffusion coefficient. The quantum contribution to $D$ reads
\begin{equation}
\label{dD:full}
\delta D = -\frac{2}{Mk_B T} \int_0^\infty d\varepsilon \exp{\left(-\frac{\varepsilon}{k_B T}\right)} D(\varepsilon) C(\varepsilon),
\end{equation}
where 
$D(\varepsilon) = (\varepsilon/M)\tau$ is the energy-dependent classical diffusion coefficient and $C(\varepsilon)$ represents the sum of the maximally crossed diagrams [inner part of Fig.~\ref{fig:wl}(c)]. We present $C(\varepsilon) = 2\sum_{\bm q} \int_0^\infty dt C_{\bm q}(\varepsilon;t,-t)$ with $C_{\bm q}(\varepsilon;t,t')$ being the Cooperon and following Refs.~\cite{golubentsev,afonin:1985,PhysRevA.36.5729}  derive the equation for the Cooperon in the form
\begin{equation}
\label{C}
\left[\frac{\partial}{\partial t} + D(\varepsilon) q^2 + \frac{1}{\tau_r} + \frac{1}{\tau}\Phi(t)\right]C_{\bm q}(\varepsilon;t,t')=\delta(t-t').
\end{equation}
The first three terms are standard and describe diffusive propagation of excitons in the disordered system and the last one accounts for the phase breaking due to inelasticity of the exciton-phonon interaction. The function $\Phi(t)=\varepsilon Ms^2t^2/(2\hbar^2)$~\cite{supp}. Solving Eq.~\eqref{C} and substituting the result into Eq.~\eqref{dD:full} we arrive at~\cite{supp}
\begin{equation}
\label{dD:full:1}
\delta D= -  \frac{\hbar}{2\pi M} \ln{\left(\frac{\tau_\phi}{\tau}\right)}, \quad \tau_\phi = \left(\frac{\hbar^2 \tau_0}{(k_BT)^2}\right)^{1/3}.
\end{equation}
In Eq.~\eqref{dD:full:1} we kept only logarithmically large contributions to $\delta D$ and assumed $\tau_r \gg \tau_\phi$. 
Equations~\eqref{D:ac} and \eqref{dD:full:1} in agreement with the qualitative analysis above [Eq.~\eqref{dD:est}], describe, respectively, the classical and quantum contributions to the exciton diffusion coefficient. At the dominant acoustic phonon scattering $\tau_\phi/\tau \propto T^{1/3}$.

\begin{figure}[t]
\includegraphics[width=0.99\linewidth]{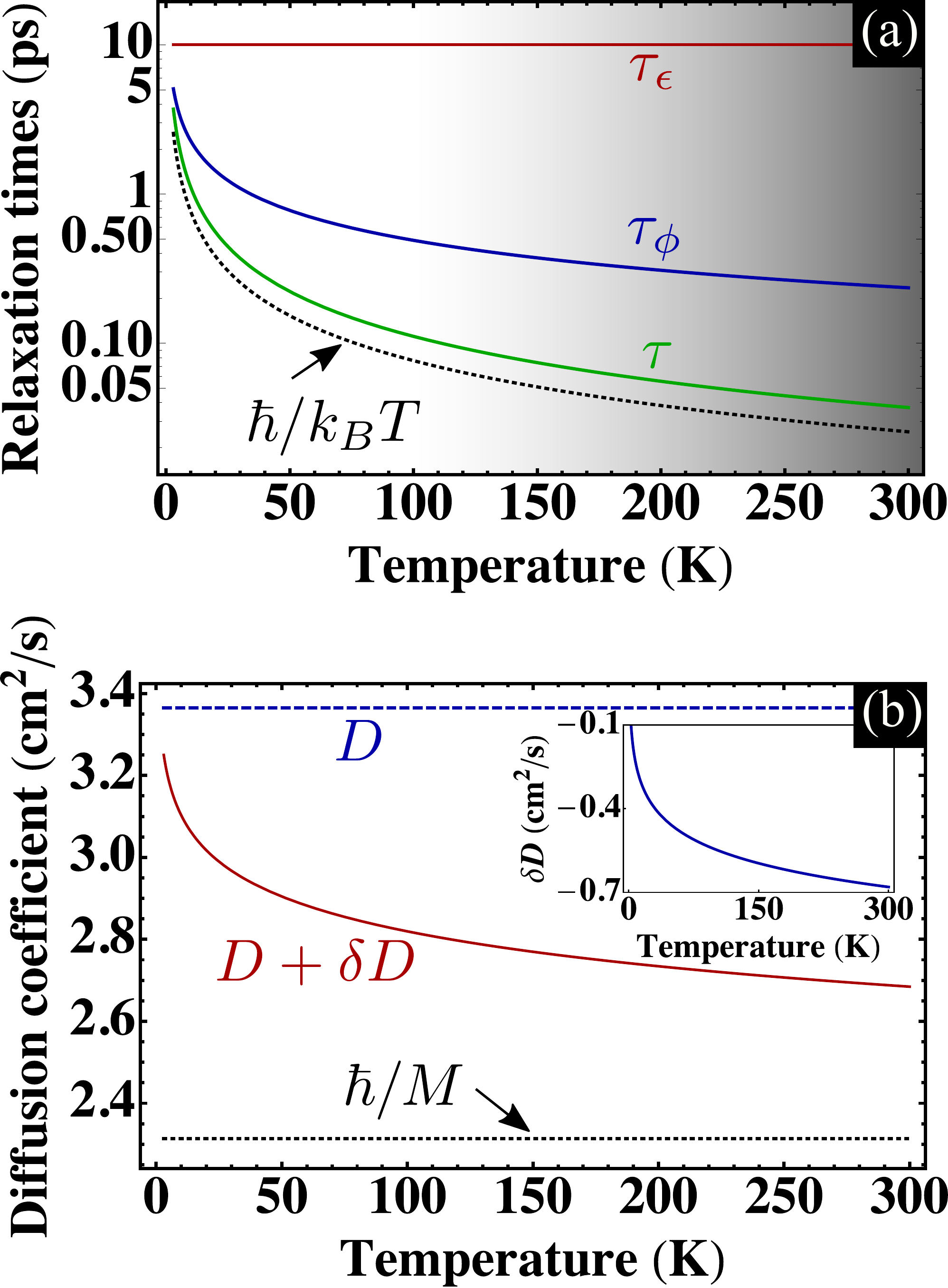}
\caption{\textbf{Effect of acoustic phonons.} (a) Exciton scattering times as functions of temperature calculated for the parameters typical for TMDC MLs: deformation potential difference $|\Xi_c - \Xi_v|=7$~eV, exciton translational mass $M=0.5m_0$ with $m_0$ being free electron mass, mass density $\rho=4.46\times 10^{-7}$~g/cm$^2$, speed of sound $s=4.1\times 10^5$~cm/s~\cite{RevModPhys.90.021001,PhysRevB.85.115317,PhysRevB.90.045422,shree2018exciton,Phuc:2018aa}. Dotted curve shows $\hbar/(k_B T)$ for comparison. In the shaded area optical and zone-edge phonon scattering can be important. (b) Classical value of the diffusion coefficient $D$ (blue) and total value of the diffusion coefficient $D+\delta D$ (red) calculated after Eqs.~\eqref{D:ac} and \eqref{dD:full:1}. Dotted curve shows limit for classical diffusion $\hbar/M$. Inset demonstrates the quantum correction $\delta D(T)$. }\label{fig:phon}
\end{figure}

\emph{Results and discussion}.
Figure~\ref{fig:phon}(a) shows the results of the calculation of the relaxation times for the excitons interacting with acoustic phonons: the energy relaxation time $\tau_\epsilon=\tau_0/2$, the phase relaxation time $\tau_\phi$, Eq.~\eqref{dD:full:1}, and the momentum relaxation time $\tau$, Eq.~\eqref{tau:ac} for the typical parameters of 2D TMDC (see caption for details). Due to significant spread of the deformation potential values reported in the literature on the basis of atomistic calculations, temperature-dependent exciton linewidth broadening and strain-tuning of exciton resonances~\cite{PhysRevB.85.115317,PhysRevB.90.045422,shree2018exciton,Phuc:2018aa,2053-1583-2-1-015006,2053-1583-3-2-021011} we used relatively large difference $|\Xi_c - \Xi_v|=7$~eV, which, however, provides a reasonable $60$~$\mu$eV/K temperature dependent linewidth broadening and consistent with experiments strain-induced shift of the exciton resonance. In the relevant temperature range the exciton lifetime for non-resonant excitation exceeds $10\ldots 100$~ps's and can be ignored~\cite{PhysRevB.94.205423,PhysRevB.93.205423}. The effect can be even more pronounced for the dark (intervalley and/or spin-forbidden) excitons whose lifetimes are even longer.  In agreement with qualitative analysis the longest timescale at $T\gtrsim 0.5\ldots 3$~K is given by the energy relaxation time, and $\tau \ll \tau_\phi \ll \tau_\epsilon$ in a wide range of temperatures. Note that $\tau$ is just slightly longer than the `thermal' time $\hbar/(k_B T)$. The calculated diffusion coefficient (red curve) and classical value (blue dashed line) are shown in Fig.~\ref{fig:phon}(b), main panel. The classical value of $D$ is temperature independent for the acoustic phonon scattering, the quantum correction is negative  and demonstrates sizeable variation with the temperature (see inset). This quantum correction solely controls the temperature dependence of the diffusion coefficient in a wide temperature range.

\begin{figure}[t]
\includegraphics[width=0.99\linewidth]{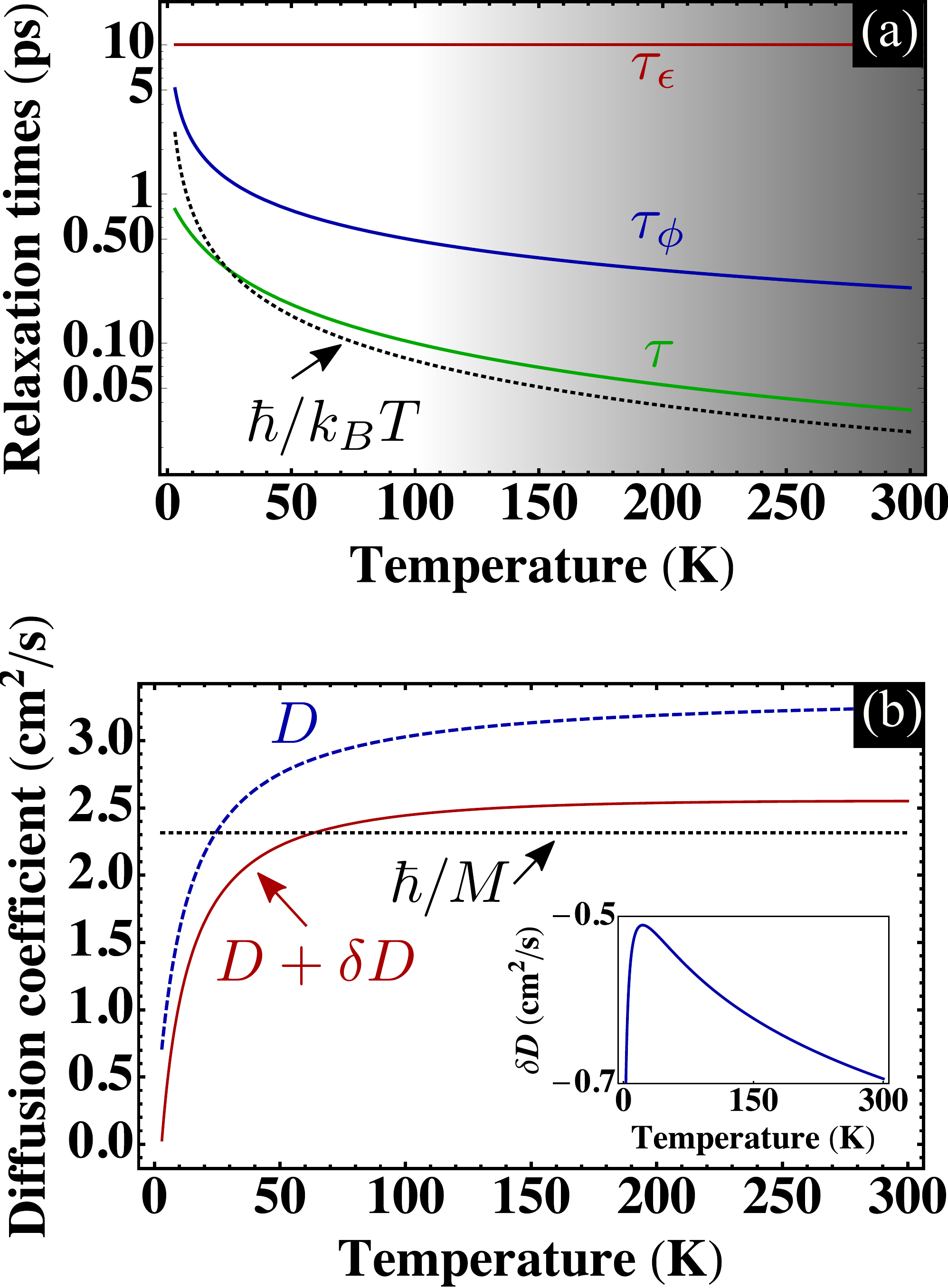}
\caption{\textbf{Effect of acoustic phonons and static disorder.} (a) Exciton scattering times as functions of temperature calculated for the same parameters as Fig.~\ref{fig:phon} also with the elastic scattering by the static disorder with $\tau_{\rm dis}=1$~ps. Dotted curve shows $\hbar/(k_B T)$ for comparison. In the shaded area optical and zone-edge phonon scattering can be important. (b)~Classical value of the diffusion coefficient $D$ (blue) and the total value of the diffusion coefficient $D+\delta D$ (red) calculated after Eqs.~\eqref{D:ac} and~\eqref{dD:full:1}. Dotted curve shows limit for classical diffusion $\hbar/M$. Inset demonstrates the~$\delta D(T)$.  }\label{fig:imp}
\end{figure}

The temperature dependence of the scattering rates and diffusion coefficients changes if other scattering processes are taken into account. For illustration we consider in Fig.~\ref{fig:imp} the situation where, in addition to the acoustic phonon scattering, the excitons interact with the short-range potential disorder characterized by the scattering time $\tau_{\rm dis}$. Here the classical diffusion coefficient already shows the temperature dependence with $D\propto T$ at low temperatures where $\tau_{\rm dis} \lesssim \tau$ and saturates with the increasing the temperature~\footnote{Due to relatively short value of $\tau_{\rm dis}$ at $T\lesssim 50$~K the criterion~\eqref{IR} is strongly violated, excitons can be localized, and the hopping transport can be realized.}. The quantum contribution $\delta D$ is now a non-monotonic function of temperature, see inset in Fig.~\ref{fig:imp}(b). Overall, the quantum correction to the diffusion coefficient is particularly pronounced due to a smaller value of the classical diffusion coefficient.

Deviations from the behavior shown in Figs.~\ref{fig:phon} and \ref{fig:imp} are expected for the temperatures $T\gtrsim 100$~K (shaded area) where the optical and zone-edge phonons come into play. Such phonons are largely dispersionless and the scattering of excitons by these phonons is inelastic. Thus three times $\tau_\epsilon$, $\tau_\phi$ and $\tau$ are of the same order of magnitude~\cite{gantmakher87}. Also the intervalley scattering by short-wavelength phonons suppresses the interference. The correction~\eqref{dD:full:1} is thus suppressed. However, the condition~\eqref{IR} of the quasi-classical transport can be violated in this case as well. The polaron effects resulting from the strong coupling of excitons with  the phonons~\cite{PhysRev.133.A1070,Firsov:2007aa,PhysRevLett.123.076601} should be  taken into account while analyzing the exciton transport at elevated temperatures.

We also briefly address the role of exciton-exciton interactions in the diffusion. While the scattering between identical particles does not affect the diffusion~\footnote{This is not generally the case where several non-equivalent valleys with different exciton masses are occupied. Such behavior will be analyzed elsewhere.} exciton-exciton scattering breaks the phase of the diffusing particle and contributes additively to $\tau_\phi^{-1}$. Unlike the dephasing caused by the electron-electron interaction~\cite{aa,Altshuler_1982,Aleiner:1999aa} the exciton-exciton scattering in MLs is mainly controlled by the exchange interaction~\cite{PhysRevB.96.115409} and is strongly inelastic. The dephasing rate can be estimated for non-degenerate excitons as $\tau_{\phi,xx}^{-1} \sim (na_B^2) E_B/\hbar$, where $n$ is the exciton density, $a_B$ and $E_B$ are the exciton Bohr radius and binding energy, respectively. For $n=10^{10}$~cm$^{-2}$, $E_B=200$~meV and $a_B=15$~\AA, the $\tau_{\phi,xx} \sim 10$~ps and exceeds phonon-induced dephasing rate. This effect, however, can be controlled by the variation of photogenerated exciton density.

Importantly, it follows from Eqs.~\eqref{D}, \eqref{IR}, and \eqref{D:ac} that the classical value of the diffusion coefficient cannot potentially be lower than $D=\hbar/M \sim $~cm$^2$/s regardless the scattering mechanism and the temperature. This limit is shown in Figs.~\ref{fig:phon}(b) and \ref{fig:imp}(b) by dotted lines. Hence, the values of the room temperature exciton diffusion coefficient reported  in Ref.~\cite{PhysRevLett.120.207401} for non-encapsulated samples $D\approx 0.3$~cm$^2$/s  require the description beyond the quasi-classical approach including additionally the effects of strong localization and hopping, and, potentially,  polarons. Encapsulation in a BN results in an enhancement of the room temperature exciton diffusion coefficient up to $1\ldots 10$ of cm$^2$/s~\cite{Cadiz:2018aa,zipfel2019exciton,Raja:2019aa} somewhat reducing the relative role of quantum effects. Systematic studies of temperature dependence of the exciton diffusion coefficient are anticipated to fully clarify the role of the quantum interference in the exciton transport in 2D TMDC.

\emph{Conclusion}.
We have demonstrated theoretically the importance of quantum interference in exciton transport. The weak localization decreases the exciton diffusion coefficient as compared to the classically calculated value. For the relevant case of the exciton interaction with acoustic phonons the temperature dependence of the diffusion coefficient is solely provided by the quantum interference. Calculations show that the effect is easily accessible for monolayer and bilayer transition metal dichalcogenides making them prospective for  studies of quantum transport physics in excitonic systems.

\begin{acknowledgements}
Valuable discussions with L.E. Golub, A. Chernikov, and E. Malic are gratefully acknowledged. 
This work has been partially supported by RFBR Project \# 19-02-00095.
\end{acknowledgements}

\end{document}

\newcount\timehh  \newcount\timemm
\timehh=\time \divide\timehh by 60
\timemm=\time
\count255=\timehh\multiply\count255 by -60 \advance\timemm by \count255

\title{Supplementary information to\\
Quantum interference effect on exciton transport in monolayer semiconductors}

\author{M.M. Glazov}

\address{Ioffe Institute, 194021 St.-Petersburg, Russia}

\maketitle

\tableofcontents

\section{Correlation function of the phonon-induced potential}\label{sec:corr}

The exciton-phonon scattering can be considered as a result of the exciton interaction with effective potential $U(\bm r, t)$ fluctuating in the space, $\bm r$, and time, $t$. We introduce the correlation function of the potential
\begin{equation}
\label{corr:UU}
\mathcal K(\bm r_1 - \bm r_2, t_1 - t_2) = \langle U(\bm r_1, t) U(\bm r_2, t)\rangle.
\end{equation}
Note that our system is the translationally invariant, that is why the correlation function $\mathcal K$ depends only the coordinate and time differences. The averaging in Eq.~\eqref{corr:UU} is carried out over the phonon ensemble. It is convenient to introduce the Fourier image of the correlator
\begin{equation}
\label{corr:UU:F}
\mathcal K_{\bm q,\omega} = \int \mathcal K(\bm r, t) e^{-\mathrm i \bm q \bm r +\mathrm i\omega t}\, d\bm r \, dt .
\end{equation}
It can be readily evaluated via the explicit expressions for the exciton-phonon interaction matrix elements~[see, e.g., \cite{shree2018exciton}]
\begin{equation}
\label{matrix:elements}
U(\bm r, t) = \sum_{\bm q} \sqrt{\frac{\hbar}{2\rho \Omega_q \mathcal S}} q b_{\bm q}^\dag(\Xi_c -\Xi_v) e^{-\mathrm i \bm q \bm r+ \mathrm i \Omega_q t}+{\rm c.c.},
\end{equation}
Here $\bm q$ is the phonon wavevector, $\Omega_q = s q$ with $s$ being the speed of sound is its dispersion, $\Xi_c$ and $\Xi_v$ are, respectively, the conduction and valence band deformation potentials, $b_{\bm q}^\dag$ ($b_{\bm q}$) are the phonon creation (annihilation) operators, $\mathcal S$ is the normalization area.
Taking into account that the phonons with different wavevectors are independent we arrive at 
\begin{multline}
\label{corr:UU:F:1}
\mathcal K_{\bm q,\omega} = \frac{\pi  k_B T}{\rho s^2 \mathcal S} (\Xi_c -\Xi_v)^2 
[\delta(\omega - \Omega_q) +  \delta(\omega + \Omega_q) ].
\end{multline}
Here we used the approximate form of the phonon distribution function
\begin{equation}
\label{Planck}
\langle b_{\bm q}^\dag b_{\bm q} \rangle = \langle b_{\bm q} b_{\bm q}^\dag \rangle -1 = n(\Omega_q) \approx \frac{k_B T}{\hbar s q} \gg 1,
\end{equation}
at $\hbar\Omega_q \ll k_B T$; these phonons are relevant for the transport phenomena at not too low temperatures, see the main text for details.

Equation~\eqref{corr:UU:F:1} allows us to evaluate the exciton momentum relaxation time relevant for the calculation of the classical value of the diffusion coefficient. Making use of the Fermi golden rule we arrive at
\begin{equation}
\label{momentum}
\frac{1}{\tau(E_p)} = \frac{1}{\hbar} \sum_{\bm p'} (1-\cos{\phi_{\bm p,\bm p'}})\mathcal K_{\bm p -\bm p', \omega_{\bm p-\bm p'}}.
\end{equation}
Here $\bm p$ and $\bm p'$ are the exciton momenta before and after the scattering, $\phi_{\bm p,\bm p'}$ is the angle between the initial and final momenta, 
\[
\omega_{\bm p-\bm p'} = \frac{E_p  - E_{p'}}{\hbar}
\]
is the transferred frequency, $E_p= p^2/(2M)$ is the exciton dispersion. At $\hbar\Omega_q \ll k_B T$ (this condition is satisfied at not too low temperatures $k_B T \gg Ms^2$) the collisions are quasi-elastic and we obtain
\begin{equation}
\label{momentum:1}
\frac{1}{\tau} = \frac{k_B T}{Ms^2} \frac{(\Xi_c -\Xi_v)^2 M^2}{\rho s^2 \hbar^3},
\end{equation}
in agreement with the main text. We note that the scattering is isotropic and the out-scattering rate equals to the momentum scattering rate.

It follows from Eq.~\eqref{corr:UU:F:1} that the Fourier-transform of the correlation function depends on the wavevector only via the phonon dispersion in the $\delta$-functions. Furthermore, for the relevant frequencies $\omega \sim k_B T/\hbar$ the $\Omega_q$ in Eq.~\eqref{corr:UU:F:1} can be set to zero. As a result, the correlator
\begin{equation}
\label{corr:UU:r}
\mathcal K(\bm r_1  - \bm r_2, t_1 - t_2) = \frac{ k_B T}{\rho s^2} (\Xi_c -\Xi_v)^2 \delta(\bm r_1 - \bm r_2).
\end{equation}
Such form of the correlator is typical for the static disorder. It is short-range in the real space and the effective potential is time-independent. 

The allowance for the finite phonon frequencies gives rise to the dephasing, see Sec.~\ref{sec:prop}. In the lowest approximation in the small parameter $\hbar\Omega_q/(k_B T) \ll 1$ we have 
\begin{equation}
\label{corr:UU:r:1}
\mathcal K(\bm r, t) = \frac{ k_B T}{\rho s^2} (\Xi_c -\Xi_v)^2 \int_0^\infty J_0(qr) \cos{(sqt)} \frac{q dq}{2\pi},
\end{equation}
with $J_0(x)$ being the Bessel function of the first kind. The integral demonstrates that the phonon-induced potential fluctuations propagate with the speed of sound.

\section{Details of the derivations of quantum corrections to the diffusion coefficient}\label{sec:deriv}

The retarded, $G^R_{\bm p}(\varepsilon)$ and advanced $G^{A}_{\bm p}(\varepsilon)$ Greens functions of excitons can be evaluated within the Born approximation neglecting quasi-elasticity of the scattering. The Greens functions take the form
\begin{equation}
\label{greens}
G^{R/A}_{\bm p}(\varepsilon) =\left(\varepsilon - \frac{\hbar^2 p^2}{2M} \pm \mathrm i \frac{\hbar}{2\tilde \tau}\right)^{-1}.
\end{equation}
Here $\tilde\tau^{-1} = \tau^{-1} + \tau_r^{-1}$ with $\tau_r$ being the exciton lifetime and  $\tau$ being the momentum relaxation time, Eq.~\eqref{momentum:1} and Eq. (3) of the main text. The account for the non-elasticity  only slightly changes $\tilde \tau$ in the denominator.

The quantum interference contribution to $D$ reads resulting from the maximally crossed diagrams  in Fig.~1(c) of the main text can be expressed in the following form
\begin{equation}
\label{dD:full}
\delta D = -\frac{2}{Mk_B T} \int_0^\infty d\varepsilon \exp{\left(-\frac{\varepsilon}{k_B T}\right)} D(\varepsilon) C(\varepsilon),
\end{equation}
where 
$D(\varepsilon) = (\varepsilon/M)\tau$ is the energy-dependent classical diffusion coefficient and $C(\varepsilon)$ represents the sum of the maximally crossed diagrams [inner part of Fig.~1(c) of the main text]. 

We make standard transformations and present 
\[
C(\varepsilon) = 2\sum_{\bm q} \int_0^\infty dt C_{\bm q}(\varepsilon;t,-t),
\]
with $C_{\bm q}(\varepsilon;t,t')$ being the Cooperon. We  follow Refs.~\cite{golubentsev,afonin:1985,PhysRevA.36.5729}  and derive the equation for the Cooperon in the form
\begin{equation}
\label{C}
\left[\frac{\partial}{\partial t} + D(\varepsilon) q^2 + \frac{1}{\tau_r} + \frac{1}{\tau}\Phi(t)\right]C_{\bm q}(\varepsilon;t,t')=\delta(t-t').
\end{equation}
The first three terms are standard and describe diffusive propagation of excitons in the disordered system and the last one accounts for the phase breaking due to inelasticity of the exciton-phonon interaction. The function 
\[
\Phi(t)= 1-\left\langle\cos{\left({\omega_{\bm p - \bm p'}t}/{2}\right)}\right\rangle,
\]
where the averaging $\langle\ldots\rangle$ is carried out over the angle between the exciton momenta $\bm p$ and $\bm p'$ before and after the scattering with the absolute values $p=p'=\sqrt{2M\varepsilon}$, $\hbar\omega = s|\bm p - \bm p'|$ is the transferred energy. In order to study the dephasing it is sufficient to decompose $\Phi(t)$ up to the $t^2$ term with the result
\[
\Phi(t) = \frac{\varepsilon Ms^2  t^2}{2\hbar^2}.
\] 
As a result,
\begin{equation}
\label{C:sol}
C_{\bm q}(\varepsilon;t,-t) = \Theta(t)\exp{\left[-2D(\varepsilon)q^2t - \frac{t}{\tau_r} -\frac{2t^3}{\tau^3_\phi(\varepsilon)} \right]},
\end{equation}
where the energy-dependent phase relaxation time $\tau_\phi(\varepsilon) = [{Ms^2\varepsilon}/({6\hbar^2\tau})]^{-1/3}$. The peculiar form of the dephasing $\propto (t/\tau_\phi)^3$ results from the quasi-elasticity of the collisions with phonons~\cite{PhysRevB.36.5663,afonin:1985}, see also Sec.~\ref{sec:prop}. 
Finally we arrive at
\begin{equation}
\label{dD:full:1}
\delta D= -  \frac{\hbar}{2\pi M} \ln{\left(\frac{\tau_\phi}{\tau}\right)}, \quad \tau_\phi = \left(\frac{\hbar^2 \tau_0}{(k_BT)^2}\right)^{1/3}.
\end{equation}
In Eq.~\eqref{dD:full:1} we kept only logarithmically large contributions to $\delta D$ and assumed $\tau_r \gg \tau_\phi$.

\section{Phase breaking due to the phonon propagation}\label{sec:prop}

Quasi-elastic exciton scattering by acoustic phonons results in a specific dephasing mechanism characterized by the $\propto \exp{(-2t^3/\tau_\phi^3)}$ decay of the Cooperon at the long-time limit, Eq.~\eqref{C:sol}. It is instructive to present a qualitative description of the effect.

To that end we follow Refs.~\cite{golubentsev,PhysRevB.36.5663} and calculate explicitly the phase difference acquired by the exciton while passing the loop clock- and counterclockwise. We assume that the loop is passed clockwise during the time interval from $-t$ to $t$ and the scattering events took place at the time moments $t_i\in(-t,t)$ at the coordinates $\bm r(t_i)$. At the time-reversed propagation the scattering events take place at $-t_i$ time moments at the coordinates $\bm r(-t_i)$, respectively. The phase factor describing the interference reads
\begin{equation}
\label{C:phi}
C_\phi = \prod_i \exp{\left\{\mathrm i \frac{\Delta \bm p_i}{\hbar}[\bm r(t_i) - \bm r(-t_i)]\right\}}.
\end{equation}
Here $\Delta \bm p_i$ is the change of the exciton momentum at $i$th scattering. The non-zero difference $\bm r(t_i) - \bm r(-t_i)$ is due to the motion of phonons. Note that if the phonon-induced potential field were static, Eq.~\eqref{corr:UU:r}, the $C_\phi\equiv 1$.

We take into account the phonon propagation during the scattering. Assuming that the typical loop contains many collisions we transform the product in Eq.~\eqref{C:phi} to the exponent of the sum. Performing averaging over the trajectories by virtue of $\langle \exp{(\mathrm i \phi)}\rangle = \exp{(-\langle\phi^2\rangle/2)}$ and transforming the sum into the integral we obtain (up to the numerical factor in the exponent):
\begin{equation}
\label{C:phi:1}
C_\phi \sim \exp{\left\{ - \frac{M\epsilon^2}{\hbar^2 \tau} \int_{-t}^t [\bm r(t') - \bm r(-t')]^2 dt'\right\}}.
\end{equation}
The phonon displacement over the time $t'$ is $r(t')\sim s t'$, thus 
\begin{equation}
\label{C:phi:1}
C_\phi \sim \exp{\left\{ - \frac{M s^2 \epsilon^2}{\hbar^2 \tau} \int_{-t}^t (t')^2 dt'\right\}} \sim \exp{\left[-\frac{2t^3}{\tau^3_\phi(\varepsilon)}\right]},
\end{equation}
in agreement with Eq.~\eqref{C:sol}.